\documentclass[conference]{IEEEtran}

\usepackage{graphicx,color}   
\usepackage[cmex10]{amsmath}  
\usepackage{amssymb}
\usepackage{cite}  
\usepackage[caption=false,font=footnotesize]{subfig} 
\usepackage[utf8]{inputenc}
\usepackage[T1]{fontenc}

\newcommand{\gap}{\vspace{-1ex}}

\newtheorem{theorem}{Theorem}
\newtheorem{remark}{Remark}

\begin{document}

\title{On Achievable Rate Regions of the\\ Asymmetric AWGN Two-Way Relay Channel}
\author{\IEEEauthorblockN{Lawrence Ong, Christopher M. Kellett, and Sarah J. Johnson}
\IEEEauthorblockA{School of Electrical Engineering and
Computer Science, The University of Newcastle, Australia\\
Email: lawrence.ong@cantab.net, \{chris.kellett, sarah.johnson\}@newcastle.edu.au)}}

\maketitle

\begin{abstract}
This paper investigates the additive white Gaussian noise two-way relay channel, where two users exchange messages through a relay. Asymmetrical channels are considered where the users can transmit data at different rates and at different power levels. We modify and improve existing coding schemes to obtain three new achievable rate regions. Comparing four downlink-optimal coding schemes, we show that the scheme that gives the best sum-rate performance is (i) complete-decode-forward, when both users transmit at low signal-to-noise ratio (SNR); (ii) functional-decode-forward with nested lattice codes, when both users transmit at high SNR; (iii) functional-decode-forward with rate splitting and time-division multiplexing, when one user transmits at low SNR and another user at medium--high SNR. 
\end{abstract}

\IEEEpeerreviewmaketitle

\section{Introduction}
\begin{figure}[t]
\centering
\resizebox{5.5cm}{!}{
\begin{picture}(0,0)%
\includegraphics{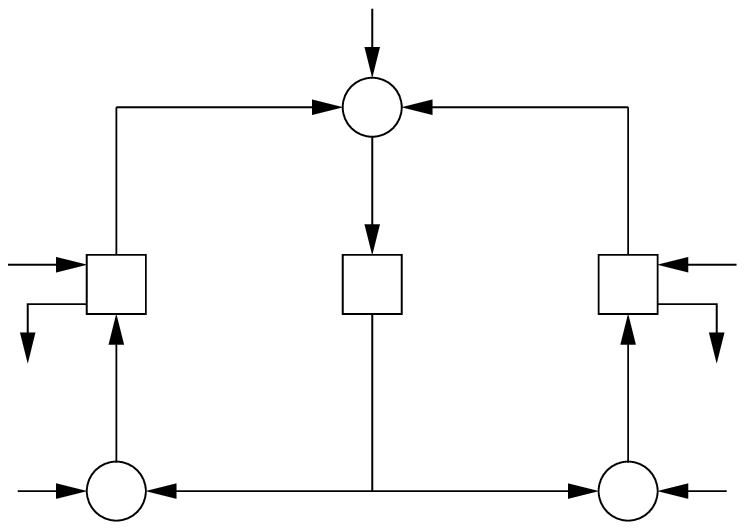}%
\end{picture}%
\setlength{\unitlength}{4144sp}%
\begingroup\makeatletter\ifx\SetFigFont\undefined%
\gdef\SetFigFont#1#2#3#4#5{%
  \fontsize{#1}{#2pt}%
  \fontfamily{#3}\fontseries{#4}\fontshape{#5}%
  \selectfont}%
\fi\endgroup%
\begin{picture}(3720,2601)(391,-2029)
\put(3871,-1501){\makebox(0,0)[lb]{\smash{{\SetFigFont{12}{14.4}{\familydefault}{\mddefault}{\updefault}{\color[rgb]{0,0,0}$\hat{W}_1$}%
}}}}
\put(3511,-961){\makebox(0,0)[lb]{\smash{{\SetFigFont{12}{14.4}{\familydefault}{\mddefault}{\updefault}{\color[rgb]{0,0,0}$2$}%
}}}}
\put(1171,-961){\makebox(0,0)[lb]{\smash{{\SetFigFont{12}{14.4}{\familydefault}{\mddefault}{\updefault}{\color[rgb]{0,0,0}$1$}%
}}}}
\put(2341,-151){\makebox(0,0)[lb]{\smash{{\SetFigFont{12}{14.4}{\familydefault}{\mddefault}{\updefault}{\color[rgb]{0,0,0}+}%
}}}}
\put(1171,-1906){\makebox(0,0)[lb]{\smash{{\SetFigFont{12}{14.4}{\familydefault}{\mddefault}{\updefault}{\color[rgb]{0,0,0}+}%
}}}}
\put(3511,-1906){\makebox(0,0)[lb]{\smash{{\SetFigFont{12}{14.4}{\familydefault}{\mddefault}{\updefault}{\color[rgb]{0,0,0}+}%
}}}}
\put(2431,-556){\makebox(0,0)[lb]{\smash{{\SetFigFont{12}{14.4}{\familydefault}{\mddefault}{\updefault}{\color[rgb]{0,0,0}$Y_3$}%
}}}}
\put(3241,-1411){\makebox(0,0)[lb]{\smash{{\SetFigFont{12}{14.4}{\familydefault}{\mddefault}{\updefault}{\color[rgb]{0,0,0}$Y_2$}%
}}}}
\put(1261,-1411){\makebox(0,0)[lb]{\smash{{\SetFigFont{12}{14.4}{\familydefault}{\mddefault}{\updefault}{\color[rgb]{0,0,0}$Y_1$}%
}}}}
\put(4096,-916){\makebox(0,0)[lb]{\smash{{\SetFigFont{12}{14.4}{\familydefault}{\mddefault}{\updefault}{\color[rgb]{0,0,0}$W_2$}%
}}}}
\put(406,-916){\makebox(0,0)[lb]{\smash{{\SetFigFont{12}{14.4}{\familydefault}{\mddefault}{\updefault}{\color[rgb]{0,0,0}$W_1$}%
}}}}
\put(2341,-961){\makebox(0,0)[lb]{\smash{{\SetFigFont{12}{14.4}{\familydefault}{\mddefault}{\updefault}{\color[rgb]{0,0,0}$0$}%
}}}}
\put(1261,-331){\makebox(0,0)[lb]{\smash{{\SetFigFont{12}{14.4}{\familydefault}{\mddefault}{\updefault}{\color[rgb]{0,0,0}$X_1$}%
}}}}
\put(3241,-331){\makebox(0,0)[lb]{\smash{{\SetFigFont{12}{14.4}{\familydefault}{\mddefault}{\updefault}{\color[rgb]{0,0,0}$X_2$}%
}}}}
\put(2431,-1771){\makebox(0,0)[lb]{\smash{{\SetFigFont{12}{14.4}{\familydefault}{\mddefault}{\updefault}{\color[rgb]{0,0,0}$X_0$}%
}}}}
\put(2296,389){\makebox(0,0)[lb]{\smash{{\SetFigFont{12}{14.4}{\familydefault}{\mddefault}{\updefault}{\color[rgb]{0,0,0}$Z_0$}%
}}}}
\put(451,-1951){\makebox(0,0)[lb]{\smash{{\SetFigFont{12}{14.4}{\familydefault}{\mddefault}{\updefault}{\color[rgb]{0,0,0}$Z_1$}%
}}}}
\put(4051,-1951){\makebox(0,0)[lb]{\smash{{\SetFigFont{12}{14.4}{\familydefault}{\mddefault}{\updefault}{\color[rgb]{0,0,0}$Z_2$}%
}}}}
\put(721,-1501){\makebox(0,0)[lb]{\smash{{\SetFigFont{12}{14.4}{\familydefault}{\mddefault}{\updefault}{\color[rgb]{0,0,0}$\hat{W}_2$}%
}}}}
\end{picture}%

}
\caption{The AWGN TWRC, where two users (nodes 1 and 2) exchange messages ($W_1$ and $W_2$) through a relay (node 0)}
\label{fig:twrc}
\vspace{-3ex}
\end{figure}

We investigate the additive white Gaussian noise two-way relay channel (AWGN TWRC) depicted in Fig.~\ref{fig:twrc}. We modify existing coding schemes and obtain three new achievable rate regions. We compare these three modified coding schemes with an existing scheme, and show that different schemes give the best sum-rate performance for different SNR regions.

The AWGN TWRC we consider has no direct link between the users; data exchange between the users is done through a relay. The AWGN TWRC is defined by two AWGN channels: the \emph{uplink} from the users to the relay, and the \emph{downlink} from the relay to the users.

If we assume that a genie informs the relay of both users' messages, and only consider the downlink,  i.e., how the relay sends these message, 
the \emph{downlink capacity region} is known~\cite{kramershamai07,oechteringschnurr08}. However, the capacity region of the AWGN TWRC is unknown in general, and the main difficulty lies in determining the best way the relay should process its received signals on the uplink.

Knopp~\cite{knopp06} proposed two coding schemes: (i) amplify-forward\footnote{This scheme was called analog relaying in \cite{knopp06}, but is now commonly referred to as amplify-forward.} where the relay simply scales its received signals on the uplink and transmits them on the downlink, and (ii) \emph{complete-decode-forward}\footnote{This scheme was called digital relaying in \cite{knopp06}, but is now commonly referred to as decode-forward. We term this scheme CDF to differentiate it from another scheme where the relay only decodes a function of the users' messages.} (CDF) where the relay decodes both users' messages on the uplink, re-encodes and sends both messages on the downlink. Schnurr et al.~\cite{schnurroechtering07} later proposed the compress-forward scheme where the relay quantizes its received signals on the uplink, re-encodes the quantized signals, and sends them on the downlink.

CDF, where the relay removes the uplink noise, is \emph{downlink
optimal} in the sense that the downlink channel usage achieves the
downlink capacity region\footnote{Decoding both
users' messages is not always optimal for the uplink.}~\cite{oechteringschnurr08}. On the other
hand, in the amplify-forward and the compress-forward schemes, the
uplink channel noise propagates to the downlink and hence they are
not downlink optimal. In this paper, we will focus on coding schemes
that are downlink optimal.

In CDF, after the relay decodes both users' messages,
instead of sending both messages on the downlink, Kramer and Shamai~\cite{kramershamai07} showed it is also downlink optimal for the relay to transmit only a function of the messages. 



Instead of decoding the individual messages and transmitting only a function of the messages, the relay might directly decode this function on the uplink. We term this scheme \emph{functional-decode-forward} (FDF). Obviously, the function must be defined such that on the downlink, each user can decode the message of the other user from the function and its own message. In addition, the channel code must \emph{match} the uplink so that the relay can decode the codeword that carries the function of the users' messages without needing to decode the messages individually.


For the AWGN TWRC where both users transmit at the same power and at the same data rate, Narayanan et al.~\cite{narayananwilson07} proposed FDF using lattice codes\footnote{Lattice codes have been shown to achieve the capacity of the point-to-point AWGN channel~\cite{erezzamir04}.} (which are linear under the modulo-lattice operation) where the relay decodes a function, i.e., modulo-lattice summation, of the user's messages.  This scheme approaches the capacity region of the AWGN TWRC asymptotically as the SNR grows. Using this scheme, both users transmit using the same lattice code and hence at the same rate. For the asymmetrical case where the users transmit at different rates, Knopp~\cite{knopp07} proposed a rate-splitting scheme as follows. The user with the lower rate transmits its message using a lattice code. The other user splits its message, and simultaneously transmits the sum (superposition) of (i) the  first part of its message using the same lattice code, and (ii) the rest of the message using a random Gaussian code. This scheme introduces interference between the lattice codeword and the Gaussian codeword. To avoid this, Nam et al.~\cite{namchung08,namchunglee09} used nested lattice codes, where one lattice code is a subset of the other lattice code, so the users transmit at different rates using lattice codes. However, this scheme suffers when there is a large difference between the users' transmit power levels. 

The contributions of this paper are as follows:
\begin{enumerate}
 \setlength{\leftskip}{-1ex} 
\item We improve the achievable rate region of FDF with nested lattice codes proposed by Nam et al.~\cite{namchunglee09}. We note that for certain SNRs, if a user transmits at a lower (than the maximum allowable) power, the achievable rate of the other user can be increased, and the sum rate can also be increased.
\item We correct and improve the achievable rate region of FDF with rate splitting and simultaneous transmission proposed by Knopp~\cite{knopp07}.  When two users transmit using the same lattice code, and the relay decodes the modulo-lattice addition of the codewords, the achievable rate of $\frac{1}{2} \log \left( 1 + \text{SNR} \right)$ used in \cite{knopp07} is incorrect. In addition, similar to FDF with nested lattice codes, we note that using less than the maximum allowable power, the achievable sum rate can be increased.
\item We propose a coding scheme using FDF with rate splitting and time-division multiplexing, and obtain a new achievable rate region. With rate splitting, one user transmits using a lattice code, while the other user uses the same lattice code and a Gaussian code. Instead of having the users transmit all codewords simultaneously, we split the transmission of the users into two phases: in the first phase, both users transmit the lattice codewords; in the second phase, one user transmits the Gaussian codeword.
\item As these schemes have the same downlink performance---all are downlink optimal---we compare their achievable sum rates on the uplink and obtain the following:
\begin{enumerate}
\item In the low SNR region, CDF outperforms the other schemes.
\item In the high SNR region, FDF with nested lattice codes outperforms the other schemes.
\item When one user transmits at low SNR and the other user at medium-to-high SNR, FDF with rate splitting and time-division multiplexing outperforms the other schemes.
\item For all SNRs, at least one of the three schemes---(1) CDF, (2) FDF with nested lattice codes, or (3) FDF with rate splitting and time-division multiplexing---is able to outperform or match FDF with rate splitting and simultaneous transmission.
\end{enumerate}
\end{enumerate}

\section{Channel Model}

The AWGN TWRC depicted in Fig.~\ref{fig:twrc} consists of three
nodes: nodes 1 and 2 are the users, and node 0 the relay. We
define by $X_i$ the transmitted signal of node $i$, and by $Y_i$ the
received signal of node $i$. The AWGN TWRC is defined by the uplink
channel $Y_0 = X_1 + X_2 + Z_0$, and the downlink channel $Y_i =
X_0 + Z_i$, for $i \in \{1,2\}$. Each $X_i$ is subject to the power constraint $E[X_i^2] \leq P_i$,
and each $Z_i$ is independent white Gaussian
noise with power $E[Z_i^2] \leq N_i$, for $i \in \{0,1,2\}$. We say that users 1 and 2
transmit at SNRs equal to $P_1/N_0$ and $P_2/N_0$
respectively\footnote{Different SNRs here can be used to model
different channel gains from the users to the relay. Varying $N_1$
and $N_2$ can be used to achieve the same effect on the channels
from the relay to the users.}.   Consider $n$ simultaneous uplink
and downlink channel uses, in which user 1 is to send an $nR_1$-bit
message $W_1$ to user 2, and user 2 is to send an $nR_2$-bit message
$W_2$ to user 1. In the $t$-th uplink channel use, each user $i$ transmits a
function of its message and its previously received signals, i.e.,
$X_i[t] = f_{i,t}(W_i,Y_i[1],Y_i[2],\dotsc,Y_i[t-1])$, for all $t \in
\{1,2,\dotsc, n\}$ and $i \in \{1,2\}$. In the $t$-th
downlink channel use, the relay transmits a function of what it
previously received, i.e., $X_0[t] = f_{0,t}(Y_0[1],
Y_0[2],\dotsc,Y_0[t-1])$. After $n$ channel uses, user 1 produces an
estimate of $W_2$ from its received messages and its own message,
$\hat{W}_2 = g_1(W_1,Y_1[1],Y_1[2],\dotsc,Y_1[n])$. User 2 does
likewise to produce $\hat{W}_1$. The rate pair $(R_1,R_2)$ is said
to be achievable if the probability of decoding error $\Pr \{
(\hat{W}_1,\hat{W}_2) \neq (W_1,W_2)\}$ can be made as small as
desired, with a sufficiently large $n$. The capacity region is the closure
of all achievable rate pairs.

\section{Existing Results}

\subsection{Capacity Outer Bound}

We define $C(x) = \frac{1}{2}\log( 1 + x)$.
An outer bound to the capacity region of the AWGN TWRC is given as follows:

\begin{theorem}[\hspace{-0.05ex}\cite{knopp07}] Consider an AWGN TWRC, a rate pair $(R_1,R_2)$ is achievable only if
\begin{align}
R_1 &\leq \min \left\{ C(P_1/N_0), C(P_0/N_2) \right\} \label{eq:ub-1}\\
R_2 &\leq \min \left\{ C(P_2/N_0), C(P_0/N_1) \right\}. \label{eq:ub-2}
\end{align}
\end{theorem}

The above outer bound can be obtained from the cut-set bound for the general multiterminal network~\cite[p. 589]{coverthomas06}. Together, the constraints $R_1 \leq C(P_0/N_2)$ and $R_2 \leq C(P_0/N_1)$ give the downlink capacity region, 
which only depends on the downlink channel parameters $P_0$, $N_1$,
and $N_2$.


\subsection{Complete-Decode-Forward}

Using CDF, the following rate region is achievable:

\begin{theorem} Consider an AWGN TWRC. CDF achieves the rate pair $(R_1,R_2)$ if
\gap
\begin{align}
R_1 &\leq C(P_1/N_0) \label{eq:cdf-1}\\
R_2 &\leq C(P_2/N_0) \label{eq:cdf-2}\\
R_1 + R_2 &\leq C([P_1+P_2]/N_0) \label{eq:cdf-3}\\
R_1 &\leq C(P_0/N_2) \label{eq:cdf-4}\\
R_2 &\leq C(P_0/N_1). \label{eq:cdf-5}
\end{align}
\end{theorem}

The above region is obtained by porting the rate region of CDF for the half-duplex discrete memoryless TWRC~\cite{knopp06} to the full-duplex AWGN TWRC. Using CDF, the encoding and decoding on the uplink are as follows:
\begin{center}
\begin{tabular}{|l | l |}
\hline
Codelength & $n$ \\
\hline
User 1 & $\boldsymbol{U}_1(W_1)$\\
\hline
User 2 & $\boldsymbol{U}_2(W_2)$\\
\hline
Relay & decodes $W_1$ and $W_2$\\
\hline
\end{tabular}
\end{center}
where $\boldsymbol{U}_i\in \mathcal{C}_{\text{Gaussian},i}$ is the length-$n$ Gaussian codeword transmitted by user $i$, $\mathcal{C}_{\text{Gaussian},i}$ is the \emph{random Gaussian} code for user $i$ with all codeletters independently generated according to the Gaussian distribution with $E[U_i^2] = P_i$. Here and in the rest of the paper, bold letters denote vectors.

The relay decodes both $W_1$ and $W_2$ on the uplink (i.e., a multiple-access channel), and it can \emph{reliably} do so (i.e., with arbitrarily small decoding error) if \eqref{eq:cdf-1}--\eqref{eq:cdf-3} are satisfied~\cite[p. 526]{coverthomas06}. On the downlink, the relay sends $(W_1,W_2)$. Knowing its own message, each relay can reliably decode the message of the other user if \eqref{eq:cdf-4} and \eqref{eq:cdf-5} are satisfied~\cite{tuncel06}.

\section{Functional-Decode-Forward}

\subsection{FDF with Nested Lattice Codes}

We improve on the FDF with nested lattice codes scheme developed by Nam et al.  \cite{namchunglee09} where both users transmit at their maximum allowable power. We modify the scheme such that user $i$ transmits at power $\delta_iP_i$ where $0 \leq \delta_i \leq 1$, for $i \in \{1,2\}$, and obtain the following new achievable rate region.

\begin{theorem} \label{thm:fdf-lattice} Consider the AWGN TWRC. FDF with nested lattice codes achieves the rate pair $(R_1,R_2)$ if
\gap
  \begin{align}
    R_1 &\leq \left[ \frac{1}{2} \log \left( \frac{\delta_1P_1}{\delta_1P_1+\delta_2P_2} + \frac{\delta_1P_1}{N_0}\right) \right]^+ \label{eq:nested-1}\\
    R_2 &\leq \left[ \frac{1}{2} \log \left( \frac{\delta_2P_2}{\delta_1P_1+\delta_2P_2} + \frac{\delta_2P_2}{N_0}\right) \right]^+ \label{eq:nested-2}\\
R_1 &\leq C(P_0/N_2) \label{eq:nested-3}\\
R_2 &\leq C(P_0/N_1), \label{eq:nested-4}
  \end{align}
for some $0 \leq \delta_1,\delta_2 \leq 1$. Here $[x]^+ = \max \{x ,0\}$.
\end{theorem}

Note that setting $\delta_1=\delta_2=1$ might not give the largest rate region as increasing $\delta_1$ decreases the RHS of \eqref{eq:nested-2}, and increasing $\delta_2$ decreases the RHS of \eqref{eq:nested-1}.

Without loss of generality, assume that $\delta_1P_1 \geq \delta_2P_2$. Using FDF with nested lattice codes, denoted by $\mathcal{C}_{\text{lattice},1}$ and $\mathcal{C}_{\text{lattice},2}$, where $\mathcal{C}_{\text{lattice},2} \subseteq \mathcal{C}_{\text{lattice},1}$, 
the users transmit the following:
\begin{center}
\begin{tabular}{|l | l |}
\hline
Codelength& $n$ \\
\hline
User 1 & $[\boldsymbol{V}_1(W_1) + \boldsymbol{D}_1] \mod \Lambda_1$ \\
\hline
User 2 & $[\boldsymbol{V}_2(W_2) + \boldsymbol{D}_2] \mod \Lambda_2$\\
\hline
Relay & decodes the function $\boldsymbol{F} \triangleq$\\
&$[ \boldsymbol{V}_1(W_1) + \boldsymbol{V}_2(W_2) + \boldsymbol{K}] \mod \Lambda_1$\\
\hline
\end{tabular}
\end{center}
where $\boldsymbol{V}_i(W_i) \in \mathcal{C}_{\text{lattice},i}$ is the length-$n$ lattice codeword for user $i$,
the lattices $\Lambda_1$ and $\Lambda_2$ satisfy $\Lambda_1 \subseteq \Lambda_2$, $\boldsymbol{D}_1$ and $\boldsymbol{D}_2$ are  randomly generated length-$n$ dither vectors which are known to all nodes and are fixed for all transmissions, $\!\!\!\! \mod \Lambda_i$ is the modulo-lattice operation~\cite{erezzamir04}, $\boldsymbol{K}$ is a deterministic function of $(\boldsymbol{V}_2(W_2) + \boldsymbol{D}_2)$,  and $\boldsymbol{K}\!\!  \mod \Lambda_2 = 0$.
If \eqref{eq:nested-1}--\eqref{eq:nested-2} are satisfied, the relay can decode $\boldsymbol{F}$ reliably~\cite{namchunglee09}. The relay then sends $\boldsymbol{F}$ on the downlink. If \eqref{eq:nested-3} and \eqref{eq:nested-4} are satisfied, both users can reliably decode $\boldsymbol{F}$. User 1 performs $(\boldsymbol{F} - \boldsymbol{V}_1)\!\! \mod \Lambda_2$ to obtain $W_2$, and user 2 performs $(\boldsymbol{F} - \boldsymbol{V}_2 - \boldsymbol{K})\!\! \mod \Lambda_1$ to obtain $W_1$.

\subsection{FDF with Rate Splitting and Simultaneous Transmission} \label{sec:fdf-rs}

Next, we correct and modify the achievable rate region using FDF with rate splitting and simultaneous transmission proposed by Knopp~\cite{knopp07} to obtain the following new rate region:

\begin{theorem} Consider the AWGN TWRC where $P_1 \geq P_2$. FDF with rate splitting and simultaneous transmission achieves the rate pair $(R_1,R_2)$ if
\gap
\begin{align}
R_2 &\leq \frac{1}{2} \log \left( \frac{1}{2} + \frac{\eta_2P_2}{N_0 + \eta_1(P_1 - \eta_2P_2)} \right) \label{eq:corrected-knopp-1}\\
R_1 - R_2 &\leq C(\eta_1[P_1 - \eta_2P_2]/N_0)\label{eq:corrected-knopp-2}\\
R_1 &\leq C(P_0/N_2) \label{eq:corrected-knopp-3}\\
R_2 &\leq C(P_0/N_1),\label{eq:corrected-knopp-4}
\end{align}
for some $0 \leq \eta_1, \eta_2 \leq 1$.
\end{theorem}

Without loss of generality, assume that $R_1 > R_2$. Let $W_1 = [W_{1a}, W_{1b}]$ where $W_{1a}$ contains $nR_2$ bits, and $W_{1b}$ contains $n(R_1-R_2)$ bits. Two codes are generated: (i) a lattice code $\mathcal{C}_\text{lattice}$ and (ii) a random Gaussian code $\mathcal{C}_\text{Gaussian}$. The uplink transmissions are as follows:
\begin{center}
\begin{tabular}{|l | l |}
\hline
Codelength & $n$ \\
\hline
User 1 & $[\boldsymbol{V}(W_{1a}) + \boldsymbol{D}_1] \mod \Lambda + \boldsymbol{U}(W_{1b})$ \\
\hline
User 2 & $[ \boldsymbol{V}(W_2) + \boldsymbol{D}_2] \mod \Lambda$\\
\hline
Relay & decodes $\boldsymbol{G} \triangleq [\boldsymbol{V}(W_{1a}) + \boldsymbol{V}(W_2)] \mod \Lambda$ \\
& and then decodes $\boldsymbol{U}(W_{1b})$\\
\hline
\end{tabular}
\end{center}
where $\boldsymbol{V}(W_{1a}), \boldsymbol{V}(W_2) \in \mathcal{C}_\text{lattice}$ are length-$n$ lattice codewords from the same lattice code, $\boldsymbol{U}(W_{1b}) \in \mathcal{C}_\text{Gaussian}$ is the length-$n$ Gaussian codeword,  and $\boldsymbol{D}_1$ and $\boldsymbol{D}_2$ are randomly generated length-$n$ dither vectors which are known to all nodes and are fixed for all transmissions.

The relay first decodes $\boldsymbol{G}$ by treating $\boldsymbol{U}$ as noise, subtracts $\boldsymbol{G}$ off its received signals, and then decodes $\boldsymbol{U}$. The relay then sends $(\boldsymbol{G},\boldsymbol{U})$ on the downlink.
The above scheme was proposed in \cite{knopp07}. We make the following modifications:
\begin{itemize}
\setlength{\itemsep}{0pt} \setlength{\parskip}{0pt} \setlength{\leftskip}{-1em}
\item We note that the users might not transmit at their full available power, as the power used by user 1 to transmit $\boldsymbol{U}$ acts as an interference when the relay decodes $\boldsymbol{G}$. We propose that both users use $\eta_2P_2$ to transmit $\boldsymbol{V}$. User 1 then uses a fraction $\eta_1$ of its remaining power of $(P_1 - \eta_2P_2)$ to transmit $\boldsymbol{U}$.
\item We correct a minor error in the rate region for $R_2$ reported in \cite{knopp07} (c.f. \eqref{eq:corrected-knopp-1}). $R_2$ is the rate of the lattice code used by both users with the same power. The relay attempts to decode the modulo-lattice sum of the lattice codewords, i.e., $\boldsymbol{G}$, in the presence of channel noise of power $N_0$ and $\boldsymbol{U}$ of power $\eta_1(P_1 - \eta_2P_2)$. It has been shown in \cite{narayananwilson07,nazergastpar07allerton} that the relay can reliably decode the modulo sum of lattice codewords if $R_2 \leq \frac{1}{2} \log \left( \frac{1}{2} + \text{SNR} \right)$, where the $\text{SNR}$ in our modified scheme is $\eta_2P_2/(N_0+\eta_1[P_1 - \eta_2P_2])$. Narayanan et al. \cite{narayananwilson07} conjectured that the rate of $R_2 = \frac{1}{2} \log \left( 1 + \text{SNR} \right)$ (reported in \cite{knopp07}) cannot in fact be achieved.
\end{itemize}

The relay can reliably decode $\boldsymbol{G}$ if \eqref{eq:corrected-knopp-1} is satisfied. The relay then removes $\boldsymbol{G}$ and can reliably decode $\boldsymbol{U}$ if \eqref{eq:corrected-knopp-2} is satisfied. The relay then sends $(\boldsymbol{G},\boldsymbol{U})$ on the downlink. Knowing $W_{1b}$, user 1 removes $\boldsymbol{U}$ from its received signals, and it can decode $\boldsymbol{G}$ if \eqref{eq:corrected-knopp-4} is satisfied. It obtains $W_2$ from $\boldsymbol{G}$ and $W_{1a}$. If \eqref{eq:corrected-knopp-3} is satisfied, user 2 can decode $(\boldsymbol{G},\boldsymbol{U})$, from which it can obtain $W_1$.

\begin{remark}
It is also possible to decode the Gaussian codeword first by treating the lattice codewords as noise. However, deriving the rate expression for this scheme is difficult as the effective noise in this case is the sum of Gaussian noise and lattices. Similar difficulty is encountered when one attempts to derive the rate expression for simultaneous decoding at the relay, as lattice decoding (using ML decoding) is used to decode the modulo-lattice sum of the lattice codewords and typical set decoding is used to decode the Gaussian codeword.
\end{remark}

\begin{remark}
This uplink scheme where a user simultaneously transmits lattice and Gaussian codes was also considered by Baik and Chung~\cite{baikchung08}. However, they employed a different coding scheme on the downlink, i.e., the relay's encoding.
\end{remark}

\subsection{FDF with Rate Splitting and Time-Division Multiplexing}

Next, we propose another coding scheme by modifying the rate splitting scheme in Sec.~\ref{sec:fdf-rs}, and  obtain the following:
\begin{theorem} Consider the AWGN TWRC where $P_1 \geq P_2$. FDF with rate splitting and time-division multiplexing achieves the rate pair $(R_1,R_2)$ if
\gap
\begin{align}
R_2 &\leq \frac{\alpha}{2} \log \left( \frac{1}{2} + \frac{P_2}{\alpha N_0} \right) \label{eq:tdm-1}\\
R_1 - R_2 &\leq (1-\alpha)C\left( \frac{P_1 - P_2}{(1-\alpha)N_0} \right) \label{eq:tdm-2}\\
R_1 &\leq C(P_0/N_2) \label{eq:tdm-3}\\
R_2 &\leq C(P_0/N_1), \label{eq:tdm-4}
\end{align}
for some $0 \leq \alpha \leq 1$.
\end{theorem}

Again, we assume that $R_1 > R_2$, and we split the message $W_1 = [W_{1a},W_{1b}]$, where $W_{1a}$ has $nR_2$ bits, and $W_{1b}$ has $n(R_1-R_2)$ bits. We generate two codes: (i) a lattice code $\mathcal{C}_\text{lattice}$ with codewords of length $\alpha n$ each, and (ii) a random Gaussian code $\mathcal{C}_\text{Gaussian}$ with codewords of length $(1-\alpha)n$ each. In the first $\alpha n$ uplink channel uses, both users transmit using the same lattice code with power $P_2/\alpha$. In the next $(1-\alpha) n$ channel uses, user 1 transmits with the Gaussian code using its remaining power $(P_1 - P_2)/(1-\alpha)$. The uplink encoding and decoding are as follows:

\begin{center}
\begin{tabular}{|l | l | l|}
\hline
Codelength & $\alpha n$ & $(1-\alpha) n$ \\
\hline
User 1 & $[\boldsymbol{V}(W_{1a}) + \boldsymbol{D}_1] \mod \Lambda$ & $\boldsymbol{U}(W_{1b})$ \\
\hline
User 2 & $[\boldsymbol{V}(W_2) + \boldsymbol{D}_2] \mod \Lambda$ & $-$\\
\hline
Relay & decodes $\boldsymbol{G} \triangleq [\boldsymbol{V}(W_{1a})$ & decodes $\boldsymbol{U}(W_{1b})$\\
& $ + \boldsymbol{V}(W_2)] \mod \Lambda$ & \\
\hline
\end{tabular}
\end{center}
where $\boldsymbol{V}(W_{1a}), \boldsymbol{V}(W_2) \in \mathcal{C}_\text{lattice}$ are length-$\alpha n$ lattice codewords from the same lattice code, $\boldsymbol{U}(W_{1b}) \in \mathcal{C}_\text{Gaussian}$ is the length-$(1-\alpha)n$ Gaussian codeword, and $\boldsymbol{D}_1$ and $\boldsymbol{D}_2$ are randomly generated length-$\alpha n$ dither vectors which are known to all nodes and are fixed for all transmissions.

In the first $\alpha n$ channel uses, the relay decodes the summation of lattice codewords $\boldsymbol{G}$. The relay can reliably decode $\boldsymbol{G}$ if \eqref{eq:tdm-1} is satisfied~\cite{narayananwilson07}. The next $(1-\alpha)n$ channel uses are AWGN point-to-point channel uses from user 1 to the relay without user 2's interference. So, the relay can reliably decode $W_{1b}$ if \eqref{eq:tdm-2} is satisfied. After the relay obtains $(\boldsymbol{G},\boldsymbol{U})$, the downlink transmission is the same as that of FDF with rate splitting and simultaneous transmission. Hence, we get \eqref{eq:tdm-3}--\eqref{eq:tdm-4}.

\begin{remark}
Our proposed scheme differs from FDF with rate splitting and simultaneous transmission in (at least) the following two ways:
\begin{enumerate}
\setlength{\itemsep}{0pt} \setlength{\parskip}{0pt} \setlength{\leftskip}{-1.5ex}
\item The lattice codes and the random Gaussian codes are of different lengths. The codelengths are proportional to the time fraction of the respective transmissions.
\item There is no interference between the lattice codewords and the Gaussian codeword.
\end{enumerate}
\end{remark}

\begin{remark}
The downlink constraints on the achievable rate regions of all four schemes discussed coincide with the downlink capacity region. So, these schemes are downlink optimal.
\end{remark}

\section{Sum Rate Comparison}

In this section, we compare the sum rate $R_\text{sum} \triangleq R_1 + R_2$ of the four schemes described in the previous sections to the sum rate upper bound. Note that a larger sum rate does not necessarily mean the entire two-dimensional rate region is larger. We fix the uplink channel noise power $N_0=2$. Without loss of generality, we assume that $P_1 \geq P_2 > 0$. For the case of $P_2 \geq P_1$, we simply reverse the roles of the users.

As these four schemes are downlink optimal, we only compare their uplink performance. This can be done by setting the relay power to be sufficiently high such that the downlink constraints will always be satisfied. 

\begin{figure}[t]
\centering
\resizebox{\columnwidth}{!}{
\begin{picture}(0,0)%
\includegraphics{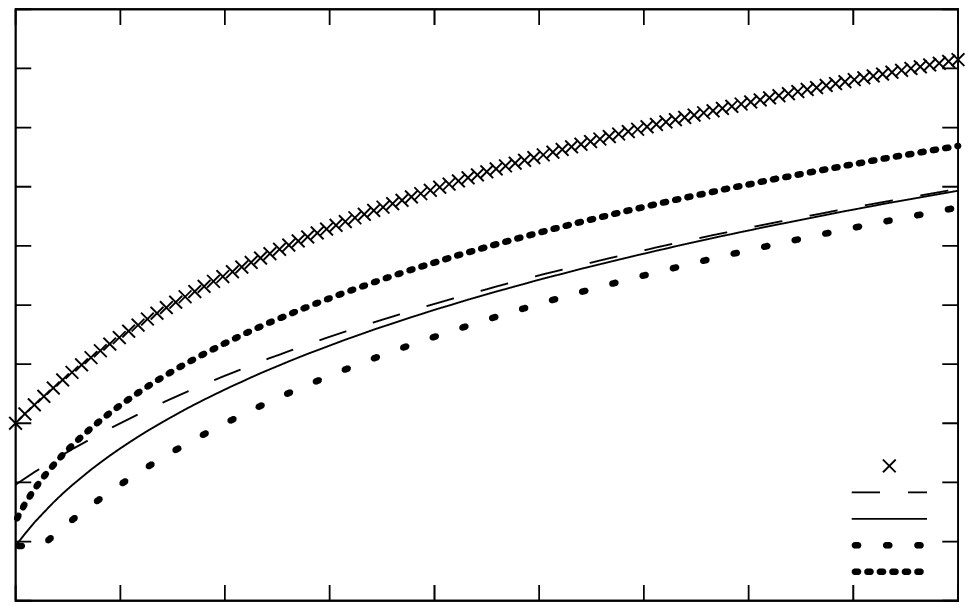}%
\end{picture}%
\setlength{\unitlength}{3947sp}%
\begingroup\makeatletter\ifx\SetFigFont\undefined%
\gdef\SetFigFont#1#2#3#4#5{%
  \reset@font\fontsize{#1}{#2pt}%
  \fontfamily{#3}\fontseries{#4}\fontshape{#5}%
  \selectfont}%
\fi\endgroup%
\begin{picture}(5271,3331)(1271,-3995)
\put(5827,-2976){\makebox(0,0)[rb]{\smash{{\SetFigFont{10}{12.0}{\familydefault}{\mddefault}{\updefault}UB}}}}
\put(5827,-3103){\makebox(0,0)[rb]{\smash{{\SetFigFont{10}{12.0}{\familydefault}{\mddefault}{\updefault}CDF}}}}
\put(5827,-3230){\makebox(0,0)[rb]{\smash{{\SetFigFont{10}{12.0}{\familydefault}{\mddefault}{\updefault}FDF (nested lattice)}}}}
\put(5827,-3357){\makebox(0,0)[rb]{\smash{{\SetFigFont{10}{12.0}{\familydefault}{\mddefault}{\updefault}FDF (rate splitting/simultaneous)}}}}
\put(5827,-3484){\makebox(0,0)[rb]{\smash{{\SetFigFont{10}{12.0}{\familydefault}{\mddefault}{\updefault}FDF (rate splitting/time division)}}}}
\put(1813,-3623){\makebox(0,0)[rb]{\smash{{\SetFigFont{10}{12.0}{\familydefault}{\mddefault}{\updefault} 0.4}}}}
\put(1813,-3339){\makebox(0,0)[rb]{\smash{{\SetFigFont{10}{12.0}{\familydefault}{\mddefault}{\updefault} 0.6}}}}
\put(1813,-3055){\makebox(0,0)[rb]{\smash{{\SetFigFont{10}{12.0}{\familydefault}{\mddefault}{\updefault} 0.8}}}}
\put(1813,-2771){\makebox(0,0)[rb]{\smash{{\SetFigFont{10}{12.0}{\familydefault}{\mddefault}{\updefault} 1}}}}
\put(1813,-2487){\makebox(0,0)[rb]{\smash{{\SetFigFont{10}{12.0}{\familydefault}{\mddefault}{\updefault} 1.2}}}}
\put(1813,-2204){\makebox(0,0)[rb]{\smash{{\SetFigFont{10}{12.0}{\familydefault}{\mddefault}{\updefault} 1.4}}}}
\put(1813,-1920){\makebox(0,0)[rb]{\smash{{\SetFigFont{10}{12.0}{\familydefault}{\mddefault}{\updefault} 1.6}}}}
\put(1813,-1636){\makebox(0,0)[rb]{\smash{{\SetFigFont{10}{12.0}{\familydefault}{\mddefault}{\updefault} 1.8}}}}
\put(1813,-1352){\makebox(0,0)[rb]{\smash{{\SetFigFont{10}{12.0}{\familydefault}{\mddefault}{\updefault} 2}}}}
\put(1813,-1068){\makebox(0,0)[rb]{\smash{{\SetFigFont{10}{12.0}{\familydefault}{\mddefault}{\updefault} 2.2}}}}
\put(1813,-784){\makebox(0,0)[rb]{\smash{{\SetFigFont{10}{12.0}{\familydefault}{\mddefault}{\updefault} 2.4}}}}
\put(1888,-3748){\makebox(0,0)[b]{\smash{{\SetFigFont{10}{12.0}{\familydefault}{\mddefault}{\updefault} 2}}}}
\put(2391,-3748){\makebox(0,0)[b]{\smash{{\SetFigFont{10}{12.0}{\familydefault}{\mddefault}{\updefault} 4}}}}
\put(2893,-3748){\makebox(0,0)[b]{\smash{{\SetFigFont{10}{12.0}{\familydefault}{\mddefault}{\updefault} 6}}}}
\put(3396,-3748){\makebox(0,0)[b]{\smash{{\SetFigFont{10}{12.0}{\familydefault}{\mddefault}{\updefault} 8}}}}
\put(3899,-3748){\makebox(0,0)[b]{\smash{{\SetFigFont{10}{12.0}{\familydefault}{\mddefault}{\updefault} 10}}}}
\put(4401,-3748){\makebox(0,0)[b]{\smash{{\SetFigFont{10}{12.0}{\familydefault}{\mddefault}{\updefault} 12}}}}
\put(4904,-3748){\makebox(0,0)[b]{\smash{{\SetFigFont{10}{12.0}{\familydefault}{\mddefault}{\updefault} 14}}}}
\put(5407,-3748){\makebox(0,0)[b]{\smash{{\SetFigFont{10}{12.0}{\familydefault}{\mddefault}{\updefault} 16}}}}
\put(5909,-3748){\makebox(0,0)[b]{\smash{{\SetFigFont{10}{12.0}{\familydefault}{\mddefault}{\updefault} 18}}}}
\put(6412,-3748){\makebox(0,0)[b]{\smash{{\SetFigFont{10}{12.0}{\familydefault}{\mddefault}{\updefault} 20}}}}
\put(1406,-2142){\rotatebox{90.0}{\makebox(0,0)[b]{\smash{{\SetFigFont{10}{12.0}{\familydefault}{\mddefault}{\updefault}$R_\text{sum}$ [bits/channel use]}}}}}
\put(4150,-3935){\makebox(0,0)[b]{\smash{{\SetFigFont{10}{12.0}{\familydefault}{\mddefault}{\updefault}$P_1$}}}}
\end{picture}%

}
\caption{Sum rate comparison, $N_0=2$, $P_2=2$}
\label{fig:compare-10}
\end{figure}
\begin{figure}[t]
\centering
\resizebox{\columnwidth}{!}{
\begin{picture}(0,0)%
\includegraphics{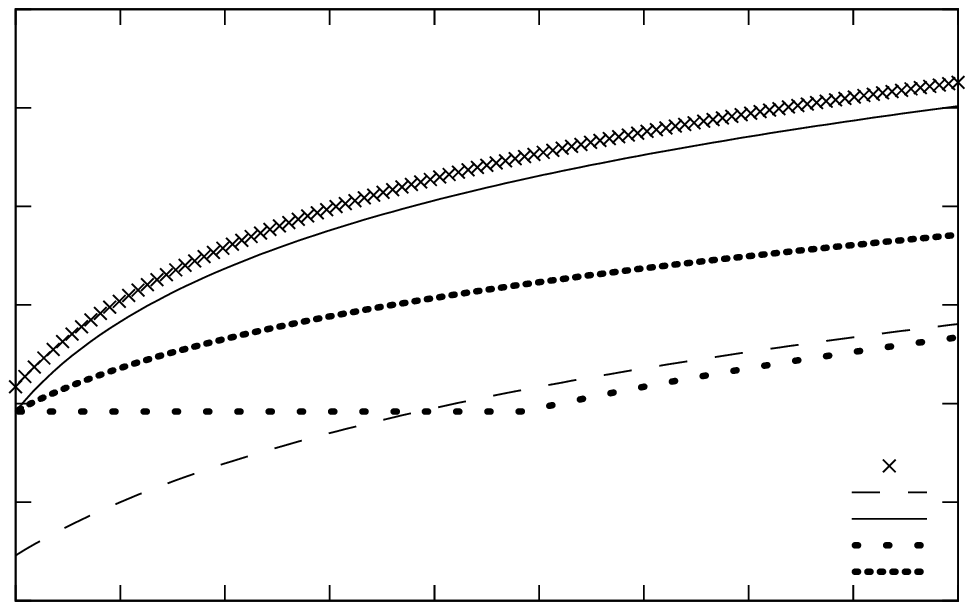}%
\end{picture}%
\setlength{\unitlength}{3947sp}%
\begingroup\makeatletter\ifx\SetFigFont\undefined%
\gdef\SetFigFont#1#2#3#4#5{%
  \reset@font\fontsize{#1}{#2pt}%
  \fontfamily{#3}\fontseries{#4}\fontshape{#5}%
  \selectfont}%
\fi\endgroup%
\begin{picture}(5317,3331)(1271,-3995)
\put(5827,-2976){\makebox(0,0)[rb]{\smash{{\SetFigFont{10}{12.0}{\familydefault}{\mddefault}{\updefault}UB}}}}
\put(5827,-3103){\makebox(0,0)[rb]{\smash{{\SetFigFont{10}{12.0}{\familydefault}{\mddefault}{\updefault}CDF}}}}
\put(5827,-3230){\makebox(0,0)[rb]{\smash{{\SetFigFont{10}{12.0}{\familydefault}{\mddefault}{\updefault}FDF (nested lattice)}}}}
\put(5827,-3357){\makebox(0,0)[rb]{\smash{{\SetFigFont{10}{12.0}{\familydefault}{\mddefault}{\updefault}FDF (rate splitting/simultaneous)}}}}
\put(5827,-3484){\makebox(0,0)[rb]{\smash{{\SetFigFont{10}{12.0}{\familydefault}{\mddefault}{\updefault}FDF (rate splitting/time division)}}}}
\put(1813,-3623){\makebox(0,0)[rb]{\smash{{\SetFigFont{10}{12.0}{\familydefault}{\mddefault}{\updefault} 1.5}}}}
\put(1813,-3150){\makebox(0,0)[rb]{\smash{{\SetFigFont{10}{12.0}{\familydefault}{\mddefault}{\updefault} 2}}}}
\put(1813,-2677){\makebox(0,0)[rb]{\smash{{\SetFigFont{10}{12.0}{\familydefault}{\mddefault}{\updefault} 2.5}}}}
\put(1813,-2203){\makebox(0,0)[rb]{\smash{{\SetFigFont{10}{12.0}{\familydefault}{\mddefault}{\updefault} 3}}}}
\put(1813,-1730){\makebox(0,0)[rb]{\smash{{\SetFigFont{10}{12.0}{\familydefault}{\mddefault}{\updefault} 3.5}}}}
\put(1813,-1257){\makebox(0,0)[rb]{\smash{{\SetFigFont{10}{12.0}{\familydefault}{\mddefault}{\updefault} 4}}}}
\put(1813,-784){\makebox(0,0)[rb]{\smash{{\SetFigFont{10}{12.0}{\familydefault}{\mddefault}{\updefault} 4.5}}}}
\put(1888,-3748){\makebox(0,0)[b]{\smash{{\SetFigFont{10}{12.0}{\familydefault}{\mddefault}{\updefault} 10}}}}
\put(2391,-3748){\makebox(0,0)[b]{\smash{{\SetFigFont{10}{12.0}{\familydefault}{\mddefault}{\updefault} 20}}}}
\put(2893,-3748){\makebox(0,0)[b]{\smash{{\SetFigFont{10}{12.0}{\familydefault}{\mddefault}{\updefault} 30}}}}
\put(3396,-3748){\makebox(0,0)[b]{\smash{{\SetFigFont{10}{12.0}{\familydefault}{\mddefault}{\updefault} 40}}}}
\put(3899,-3748){\makebox(0,0)[b]{\smash{{\SetFigFont{10}{12.0}{\familydefault}{\mddefault}{\updefault} 50}}}}
\put(4401,-3748){\makebox(0,0)[b]{\smash{{\SetFigFont{10}{12.0}{\familydefault}{\mddefault}{\updefault} 60}}}}
\put(4904,-3748){\makebox(0,0)[b]{\smash{{\SetFigFont{10}{12.0}{\familydefault}{\mddefault}{\updefault} 70}}}}
\put(5407,-3748){\makebox(0,0)[b]{\smash{{\SetFigFont{10}{12.0}{\familydefault}{\mddefault}{\updefault} 80}}}}
\put(5909,-3748){\makebox(0,0)[b]{\smash{{\SetFigFont{10}{12.0}{\familydefault}{\mddefault}{\updefault} 90}}}}
\put(6412,-3748){\makebox(0,0)[b]{\smash{{\SetFigFont{10}{12.0}{\familydefault}{\mddefault}{\updefault} 100}}}}
\put(1406,-2142){\rotatebox{90.0}{\makebox(0,0)[b]{\smash{{\SetFigFont{10}{12.0}{\familydefault}{\mddefault}{\updefault}$R_\text{sum}$ [bits/channel use]}}}}}
\put(4150,-3935){\makebox(0,0)[b]{\smash{{\SetFigFont{10}{12.0}{\familydefault}{\mddefault}{\updefault}$P_1$}}}}
\end{picture}%

}
\caption{Sum rate comparison, $N_0=2$, $P_2=10$}
\label{fig:compare-50}
\vspace{-3ex}
\end{figure}

We fix $P_2$ and plot the maximum $R_\text{sum}$ achievable by each scheme by varying $P_1$.
In Fig.~\ref{fig:compare-10}, we constrain user 2 to transmit at low SNR. When the other user (user 1) also transmits at low SNR, CDF gives the best performance. Still keeping user 2's SNR low, when user 1 transmits at high SNR, FDF with rate splitting and time-division multiplexing outperforms the other schemes. From Fig.~\ref{fig:compare-50}, when both users are transmitting at high SNR, FDF using nested lattice codes outperforms the other schemes.

\begin{figure}[t]
\centering
\resizebox{0.912\columnwidth}{!}{
\begin{picture}(0,0)%
\includegraphics{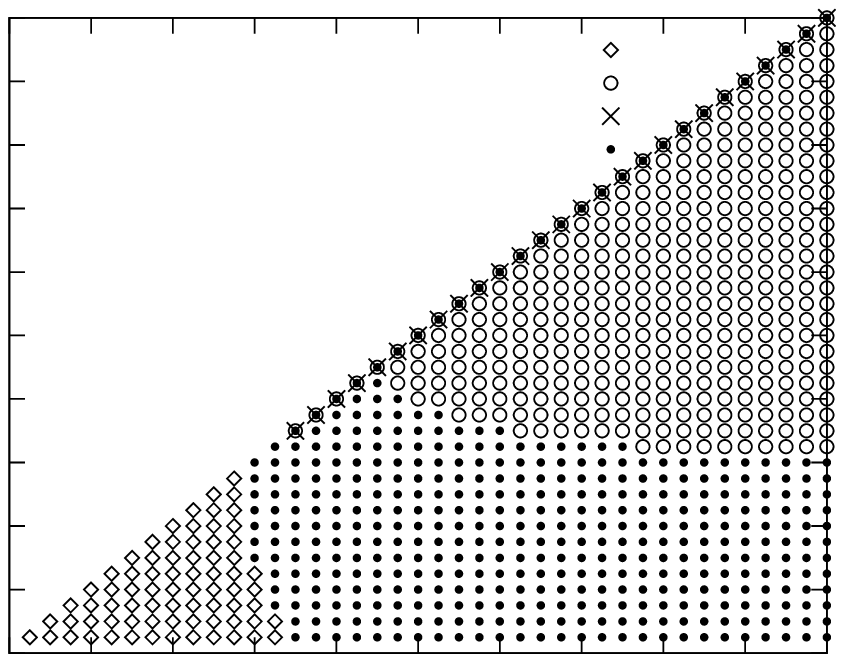}%
\end{picture}%
\setlength{\unitlength}{3947sp}%
\begingroup\makeatletter\ifx\SetFigFont\undefined%
\gdef\SetFigFont#1#2#3#4#5{%
  \reset@font\fontsize{#1}{#2pt}%
  \fontfamily{#3}\fontseries{#4}\fontshape{#5}%
  \selectfont}%
\fi\endgroup%
\begin{picture}(4612,3553)(1291,-3987)
\put(3850,-3935){\makebox(0,0)[b]{\smash{{\SetFigFont{10}{12.0}{\familydefault}{\mddefault}{\updefault}$P_1/N_0$}}}}
\put(4513,-1046){\makebox(0,0)[rb]{\smash{{\SetFigFont{10}{12.0}{\familydefault}{\mddefault}{\updefault}FDF (rate splitting/simultaneous)}}}}
\put(4513,-887){\makebox(0,0)[rb]{\smash{{\SetFigFont{10}{12.0}{\familydefault}{\mddefault}{\updefault}FDF (nested lattice)}}}}
\put(4513,-728){\makebox(0,0)[rb]{\smash{{\SetFigFont{10}{12.0}{\familydefault}{\mddefault}{\updefault}CDF}}}}
\put(1813,-3623){\makebox(0,0)[rb]{\smash{{\SetFigFont{10}{12.0}{\familydefault}{\mddefault}{\updefault} 0}}}}
\put(1813,-3318){\makebox(0,0)[rb]{\smash{{\SetFigFont{10}{12.0}{\familydefault}{\mddefault}{\updefault} 0.5}}}}
\put(1813,-3013){\makebox(0,0)[rb]{\smash{{\SetFigFont{10}{12.0}{\familydefault}{\mddefault}{\updefault} 1}}}}
\put(1813,-2708){\makebox(0,0)[rb]{\smash{{\SetFigFont{10}{12.0}{\familydefault}{\mddefault}{\updefault} 1.5}}}}
\put(1813,-2403){\makebox(0,0)[rb]{\smash{{\SetFigFont{10}{12.0}{\familydefault}{\mddefault}{\updefault} 2}}}}
\put(1813,-2098){\makebox(0,0)[rb]{\smash{{\SetFigFont{10}{12.0}{\familydefault}{\mddefault}{\updefault} 2.5}}}}
\put(1813,-1794){\makebox(0,0)[rb]{\smash{{\SetFigFont{10}{12.0}{\familydefault}{\mddefault}{\updefault} 3}}}}
\put(1813,-1489){\makebox(0,0)[rb]{\smash{{\SetFigFont{10}{12.0}{\familydefault}{\mddefault}{\updefault} 3.5}}}}
\put(1813,-1184){\makebox(0,0)[rb]{\smash{{\SetFigFont{10}{12.0}{\familydefault}{\mddefault}{\updefault} 4}}}}
\put(1813,-879){\makebox(0,0)[rb]{\smash{{\SetFigFont{10}{12.0}{\familydefault}{\mddefault}{\updefault} 4.5}}}}
\put(1813,-574){\makebox(0,0)[rb]{\smash{{\SetFigFont{10}{12.0}{\familydefault}{\mddefault}{\updefault} 5}}}}
\put(1888,-3748){\makebox(0,0)[b]{\smash{{\SetFigFont{10}{12.0}{\familydefault}{\mddefault}{\updefault} 0}}}}
\put(2280,-3748){\makebox(0,0)[b]{\smash{{\SetFigFont{10}{12.0}{\familydefault}{\mddefault}{\updefault} 0.5}}}}
\put(2673,-3748){\makebox(0,0)[b]{\smash{{\SetFigFont{10}{12.0}{\familydefault}{\mddefault}{\updefault} 1}}}}
\put(3065,-3748){\makebox(0,0)[b]{\smash{{\SetFigFont{10}{12.0}{\familydefault}{\mddefault}{\updefault} 1.5}}}}
\put(3458,-3748){\makebox(0,0)[b]{\smash{{\SetFigFont{10}{12.0}{\familydefault}{\mddefault}{\updefault} 2}}}}
\put(3850,-3748){\makebox(0,0)[b]{\smash{{\SetFigFont{10}{12.0}{\familydefault}{\mddefault}{\updefault} 2.5}}}}
\put(4242,-3748){\makebox(0,0)[b]{\smash{{\SetFigFont{10}{12.0}{\familydefault}{\mddefault}{\updefault} 3}}}}
\put(4635,-3748){\makebox(0,0)[b]{\smash{{\SetFigFont{10}{12.0}{\familydefault}{\mddefault}{\updefault} 3.5}}}}
\put(5027,-3748){\makebox(0,0)[b]{\smash{{\SetFigFont{10}{12.0}{\familydefault}{\mddefault}{\updefault} 4}}}}
\put(5420,-3748){\makebox(0,0)[b]{\smash{{\SetFigFont{10}{12.0}{\familydefault}{\mddefault}{\updefault} 4.5}}}}
\put(5812,-3748){\makebox(0,0)[b]{\smash{{\SetFigFont{10}{12.0}{\familydefault}{\mddefault}{\updefault} 5}}}}
\put(1406,-2037){\rotatebox{-270.0}{\makebox(0,0)[b]{\smash{{\SetFigFont{10}{12.0}{\familydefault}{\mddefault}{\updefault}$P_2/N_0$}}}}}
\put(4513,-1205){\makebox(0,0)[rb]{\smash{{\SetFigFont{10}{12.0}{\familydefault}{\mddefault}{\updefault}FDF (rate splitting/time division)}}}}
\end{picture}%

}
\caption{Schemes that obtain the highest sum rate for varying $P_1/N_0$ and $P_2/N_0$}
\label{fig:opt}
\vspace{-3ex}
\end{figure}

Fig.~\ref{fig:opt} gives a summary of schemes that achieve the highest sum rate for different $P_1$ and $P_2$ normalized by $N_0$.  We observe the following:
(i) When both users transmit at high SNR, FDF with nested lattice codes outperforms other schemes. (ii) When both users transmit at low SNR, CDF is the preferred scheme. (iii) When one user transmits at a low SNR and the other user at medium-to-high SNR, FDF with rate splitting and time-division multiplexing achieves the highest sum rate. In the equal-SNR region where the three FDF schemes attain the highest sum rate (i.e., $P_1/N_0=P_2/N_0 \geq 1.75$), the three schemes effectively reduce to the same scheme: both users transmit using only a lattice code (the same lattice code) at the same power and no rate splitting is done.
Also seen from the figure, FDF with rate splitting and simultaneous transmission does not give the (strictly) best sum rate at any SNR, i.e., one of the other schemes can always outperform or match it.

Using CDF, the relay needs to decode both the users' messages, c.f. \eqref{eq:cdf-3}. Because of the concavity of the $\log(\cdot)$ function, this constraint limits its performance at medium-to-high SNR. FDF, on the other hand, does not suffer from this problem. However, because of the modulo-lattice operation (see \cite{erezzamir08israel} for more discussion), FDF (which uses lattice codes) achieves rates up to $\frac{1}{2}\log(\gamma + \text{SNR})$ where $\gamma < 1$, while CDF (which uses Gaussian codes) achieves rate up to $\frac{1}{2} \log(1 + \text{SNR})$. So, FDF-based schemes do not perform well at low SNR. This explains why CDF performs better at low SNR, while FDF-based schemes perform better at medium-to-high SNR.

Using FDF with nested lattice codes, lattices of two different sizes (which depend on the transmit power) are used in the transmissions of the two users. As the relay decodes the modulo-sum of transmitted codewords with respect to the bigger lattice, the rate of the user that transmits using the smaller lattice (lower transmit power) is penalized. So, when one user transmits at high power and the other user at low power, the sum rates of FDF with nested lattice codes are affected; the sum rates of CDF are also affected by the reason given in the previous paragraph. In this region, our proposed FDF with rate splitting and time-division multiplexing is able to give better sum-rates, as it does not suffer from the problems of the need to decode both users' messages and mismatched lattices.


\bibliography{arxiv}

\end{document}